\documentclass{article}
\usepackage[T1]{fontenc} 
\usepackage[utf8]{inputenc} 
\usepackage{ismir,amsmath,cite,url}
\usepackage{graphicx}
\usepackage{color}
\usepackage{booktabs}
\usepackage{multirow}
\usepackage{comment}
\usepackage{amssymb}
\usepackage{soul}
\usepackage{todonotes}
\usepackage[normalem]{ulem}
\usepackage{amssymb}
\usepackage{pifont}
\newcommand{\cmark}{\ding{51}}%
\newcommand{\xmark}{\ding{55}}%

\usepackage{lineno}

\title{Metric learning vs classification\\for disentangled music representation learning}

\multauthor
{Jongpil Lee$^1$ \hspace{0.6cm} Nicholas J. Bryan$^2$ \hspace{0.6cm} Justin Salamon$^2$ \hspace{0.6cm} Zeyu Jin$^2$ \hspace{0.6cm} Juhan Nam$^1$
}
{
$^1$ Graduate School of Culture Technology, KAIST, Daejeon, South Korea\\
$^2$ Adobe Research, San Francisco, CA, USA\\
{\tt\small  \{richter, juhannam\}@kaist.ac.kr, \{nibryan, salamon, zejin\}@adobe.com}
}

\def\authorname{Jongpil Lee, Nicholas J. Bryan, Justin Salamon, Zeyu Jin, and Juhan Nam}

\begin{document}


%
\maketitle
\begin{abstract}
Deep representation learning offers a powerful paradigm for mapping input data onto an organized embedding space and is useful for many music information retrieval tasks. Two central methods for representation learning include deep metric learning and classification, both having the same goal of learning a representation that can generalize well across tasks. Along with generalization, the emerging concept of disentangled representations is also of great interest, where multiple semantic concepts (e.g., genre, mood, instrumentation) are learned jointly but remain separable in the learned representation space. In this paper we present a single representation learning framework that elucidates the relationship between metric learning, classification, and disentanglement in a holistic manner. For this, we (1) outline past work on the relationship between metric learning and classification, (2) extend this relationship to multi-label data by exploring three different learning approaches and their disentangled versions, and (3) evaluate all models on four tasks (training time, similarity retrieval, auto-tagging, and triplet prediction). We find that classification-based models are generally advantageous for training time, similarity retrieval, and auto-tagging, while deep metric learning exhibits better performance for triplet-prediction. Finally, we show that our proposed approach yields state-of-the-art results for music auto-tagging.

\end{abstract}

\vspace{-2mm}
\section{Introduction}\label{sec:introduction}


\begin{figure}[t!]
\vspace{-2mm}
\centering\includegraphics[width=0.7\columnwidth]{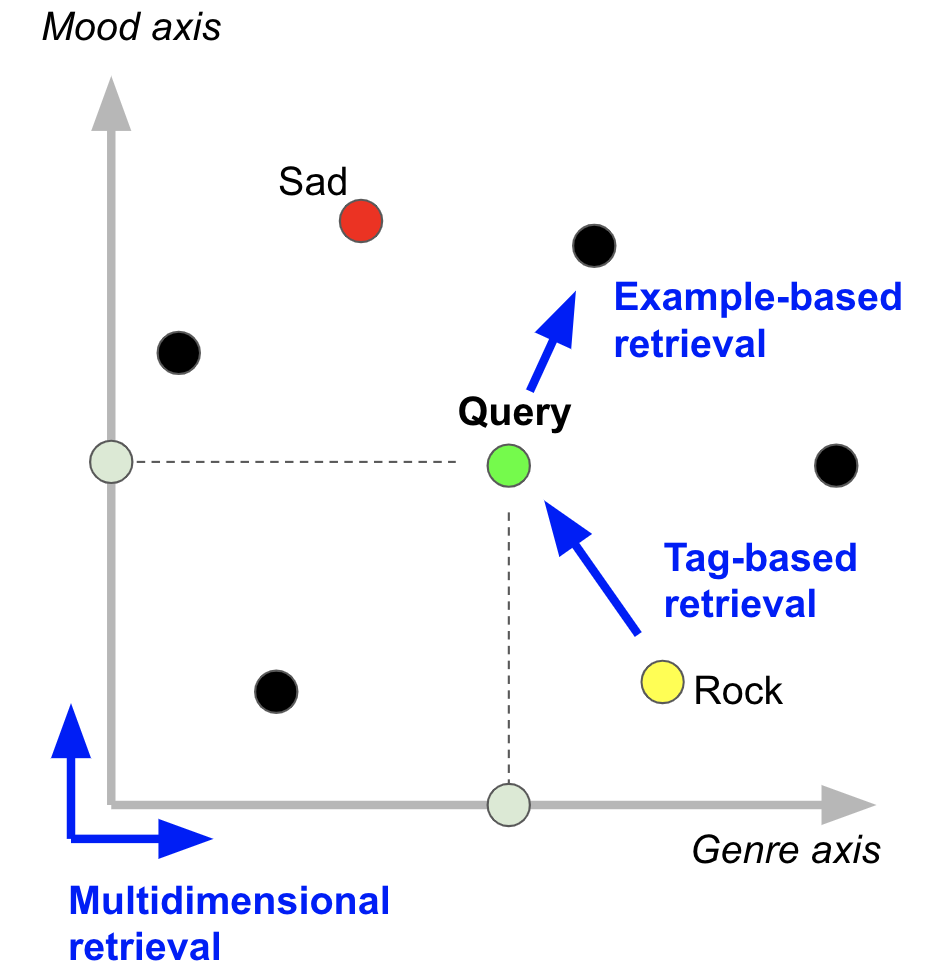}
\caption{A disentangled music representation space. The green dot depicts a query song, the black dots depict retrieval songs, the red and yellow dots depict centroids of musical concepts, the gray arrows depict multidimensional axis, and the blue arrows depict retrieval methods.}
\label{fig:figure0}
\vspace{-2mm}
\end{figure}

Learning a good representation, or embedding space, is a key goal in deep learning and is central to music classification and retrieval tasks. An important quality of a good representation is its generalization capability, i.e., its applicability to a diverse set of downstream tasks, including those relying on small datasets in a transfer learning setting~\cite{park2017representation,choi2017transfer,nam2018deep}. While numerous representation learning methods have been explored to date, two learning paradigms are particularly common: deep metric learning and classification-based representation learning. The former is based on deriving similarity scores (or distances) between examples, while the latter is achieved via a cross-entropy loss over similarity scores between example and class centroids.

While both paradigms share the goal of learning a generalizable representation, the results from each approach are generally different. For example, a learned representation optimized via a classification task may perform poorly on a similarity-search task, and vice versa. While recent studies have elucidated the theoretical relationships between these paradigms and validated them through experimental findings~\cite{zhai2018classification}, these developments have not been explored in the music domain. Furthermore, the relationship has not been explored for multi-label data, which is central to many music information retrieval tasks.

Beyond seeking a representation that generalizes across tasks, the emerging concept of disentangled representations~\cite{reed2014learning, chen2016infogan} is of great interest for music applications. Music is often labeled with multiple semantic dimensions simultaneously (e.g., genre, mood, and instrumentation) and learning a representation that can capture this structure is advantageous. We often need to search for music that is similar along a particular semantic dimension in one application (e.g., a music playlist with lighthearted mood), while requiring music similar along a different semantic dimension for another application (e.g., era for musicological analysis). Disentangled representations allow us to address both problems with a single model, and were recently proposed for audio-based music similarity search~\cite{lee2020disentangled}. However, this study only explored disentanglement via a single deep metric learning approach, and the applicability and performance of more recent metric- and classification-based learning methods is yet to be explored.


In this paper, we present a unified representation learning framework that elucidates the relationship between metric learning, classification, and disentanglement. First, we outline past work on the relationship between metric learning and classification. We then extend this relationship to multi-label and multi-concept data (common to music applications) by exploring three different learning approaches and their disentangled versions -- two of which are novel to this work. Finally, we evaluate all models against four tasks (training time, similarity retrieval, auto-tagging, and triplet prediction) and compare various aspects of the learned representations. 


\vspace{-1mm}

\section{Related Work}

\subsection{Metric Learning and Classification}
\vspace{-1mm}

The goal of distance metric learning is to obtain an embedding space where similar items are close together and dissimilar items are far apart. A common strategy is to use pairwise~\cite{chopra2005learning, hadsell2006dimensionality} or triplet-based samples to train a model~\cite{schultz2004learning, weinberger2009distance, schroff2015facenet, hoffer2015deep}. An important advantage of deep metric learning is that it can efficiently model an extremely large number of classes (e.g., for face recognition) 
\cite{schroff2015facenet}. 
However, training models using this strategy are relatively slow as models operate on triplets of input samples~\cite{movshovitz2017no}. Recently, more efficient sampling techniques have been proposed to speed up convergence, including hard negative mining, semi-hard negative mining~\cite{schroff2015facenet}, distance weighted sampling~\cite{wu2017sampling}, and proxy-based training~\cite{movshovitz2017no}.  Proxy-based training~\cite{movshovitz2017no} assigns one or several proxies to each class (given by per-class embedding centroids) and optimizes the learned space by comparing embedded input samples to proxies instead of directly comparing them to positive and negative samples. This reduces training time significantly while improving retrieval performance on images.

Classification models, on the other hand, are typically trained such that classes are linearly separable in the embedding space of the last hidden layer of the deep neural network. Since classification models are not optimized based on distances in the learned embedding space, they may not perform well when directly used for similarity-based retrieval. To overcome this, recent work proposed the application of a normalization layer over the embedding space during training, and showed that this simple technique increases model performance on similarity-based image retrieval \cite{zhai2018classification}.

Recent and parallel advances in both paradigms (metric- and classification-based learning) have shown that there is an inherent link between them~\cite{zhai2018classification, qian2019softtriple, musgrave2020metric}. The per-class embedding centroids used in proxy-based training are, in fact, equivalent to the per-class vectors obtained from the linear transformation in the last hidden layer of a classification model~\cite{qian2019softtriple}. Further, a recent comparative study demonstrated that the loss function of a triplet-based model is equivalent to that of a classification model up to a smoothing factor for single-label, multi-class data~\cite{qian2019softtriple}. These findings suggest that deep metric- and classification-based learning are not as different as initially thought and we could, potentially, use either to learn a representation that generalizes well to both similarity-based retrieval and classification tasks.

\subsection{Disentangled Representation Learning}
\vspace{-1mm}

Another important measure of representation learning is \emph{disentanglement}~\cite{ridgeway2018learning}. Recently, Lee et al.~adapted Conditional Similarity Networks (CSN) applied to triplet-based deep metric learning to the music domain \cite{veit2017conditional,lee2020disentangled}. The main idea in CSN is to apply a masking function over the embedding space, where each mask corresponds to a different semantic dimension of similarity corresponding to musical notions such as genre, mood, instrument and tempo. They showed that the disentangled music representation not only enables multidimensional music search via its sub-dimensions, but also improves general music retrieval performance when all embedding dimensions are used. However, CSN for disentangled music representation learning was only explored using a deep metric learning strategy, and classification-based approaches were not studied. Considering the close relationship between the two, we propose to study disentanglement under classification, particularly for multi-labeled music data, and compare and contrast it to disentanglement via metric learning.

\vspace{-1mm}
\section{Disentangled Learning Models}
\begin{figure*}[t!]
\vspace{-4mm}
\centering\includegraphics[width=1.0\textwidth]{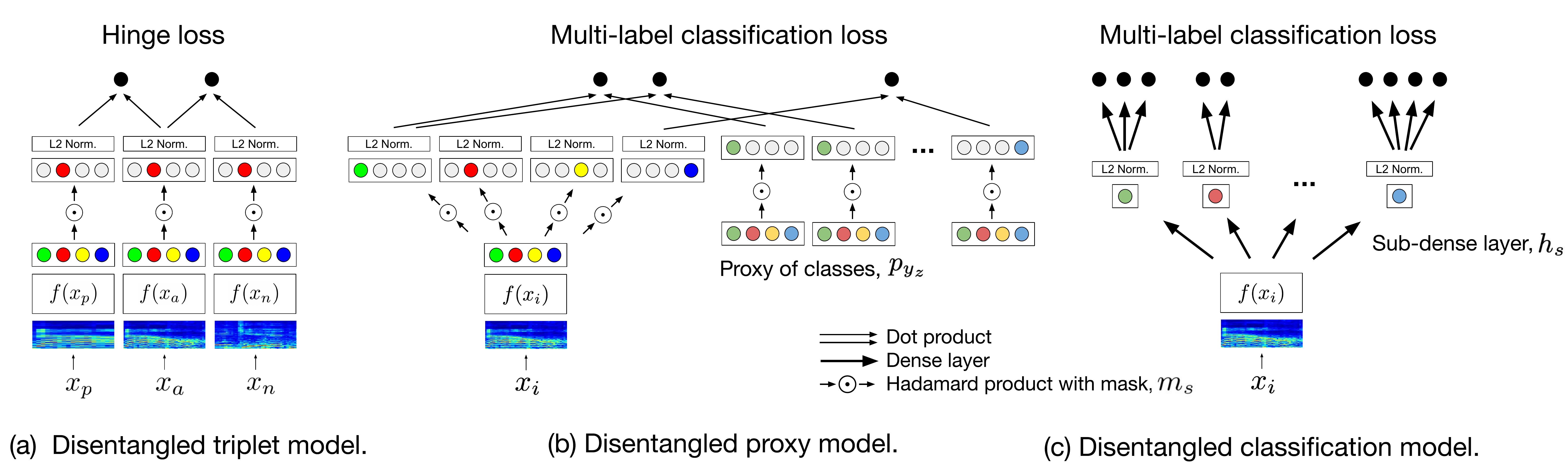}
\caption{A unified framework for disentangled triplet- and proxy-based metric learning and multi-label classification.}
\label{fig:figure1}
\vspace{-2mm}
\end{figure*}

In this section, we introduce three disentangled learning methods, which are triplet-based, proxy-based, and classification-based models. The first model was previously developed~\cite{lee2020disentangled}, and the latter two are novel contributions.
The overall architectures are illustrated in Figure \ref{fig:figure1}. In the following descriptions, $x$ denotes a data point, $f(\cdot)$ a nonlinear embedding function, $y$ a multi-hot class label, and $s$ a category (or a similarity notion such as mood, genre or instrumentation) of $y$. For example, if $y_z$ is \emph{rock}, then $s_{y_z}$ is \emph{genre}.

\subsection{Triplet-based Model}
\vspace{-1mm}

Disentangled triplet-based models were recently proposed in \cite{veit2017conditional, lee2020disentangled}. We first define a  triplet as $t=(x_a,x_p,x_n;y_z)$, where $x_a$ is the anchor sample, $x_p$ is the positive sample, and $x_n$ is the negative sample.  $x_a$ and $x_p$ are sampled to have the same positive label $y_z$, while $x_n$ is negative for $y_z$. Then, the basic triplet loss is defined as 

\vspace{-1mm}
\small
\begin{equation} \label{eq:eq1}
L(t)=\max\{0,D(f(x_a),f(x_n))-D(f(x_a),f(x_p))+\Delta\},
\end{equation}
\normalsize
where $D(f(x_i),f(x_j))=cos(f(x_i),f(x_j))$ is a distance metric, and $\Delta$ is a margin value~\cite{weinberger2009distance}. 
To disentangle the embedding feature of size $d$, a masking function $m_s\in \mathbb{R}^d$ is applied. The number of masks corresponds to the number of similarity notions $s$ and each mask occupies certain dimensions of the $\mathbb{R}^d$ space evenly as illustrated in Figure \ref{fig:figure1} (a). Thus, when the $t=(x_a,x_p,x_n;y_z)$ is used, a mask for the similarity notion $s_{y_z}$ is applied to the embedding feature space. The loss for training the model is given by: 

\vspace{-1mm}
\small
\begin{equation} \label{eq:eq2}
\begin{split}
L(t)=\max\{0,D(f(x_a)\circ m_s,f(x_n)\circ m_s) \\
-D(f(x_a)\circ m_s,f(x_p)\circ m_s)+\Delta\},
\end{split}
\end{equation}
\normalsize
where $\circ$ denotes the Hadamard product.

\subsection{Proxy-based Model}
\vspace{-1mm}

The core idea of proxy-based metric learning is that proxy embeddings are learned and assigned to each class and used to measure the distance to an anchor data point instead of directly measuring distances to pairs or triplet data samples~\cite{movshovitz2017no}. This can be interpreted as a supervised clustering algorithm, where proxies play a role of class centroids. In this approach, the distance metric becomes

\vspace{-1mm}
\small
\begin{equation} \label{eq:eq3}
D(f(x_i),p_{y_z})=cos(f(x_i),p_{y_z}) 
=\frac{f(x_i)}{||f(x_i)||}\cdot \frac{p_{y_z}}{||p_{y_z}||},
\end{equation}
\normalsize
where $x_i$ is a data point, $p_{y_z}$ is a proxy for class $y_z$, and $\cdot$ is the dot product. If the data is single-labeled (multi-class), one can apply triplet loss, Neighborhood Component Analysis (NCA) loss~\cite{goldberger2005neighbourhood}, or Softmax loss over the above distance metric \cite{movshovitz2017no, qian2019softtriple}, but with our multi-labeled data, it is not directly applicable. To address this, we replace these losses with a multi-label classification loss, i.e., binary cross entropy. The prediction score for each class becomes

\vspace{-1mm}
\small
\begin{equation} \label{eq:eq4}
\hat{y_z}=sigmoid(D(f(x_i),p_{y_z})),
\end{equation}
\normalsize
and the loss is

\vspace{-1mm}
\small
\begin{equation} \label{eq:eq5}
L(x_i)=\sum_{z}[-y_z log(\hat{y_z})-(1-y_z)log(1-\hat{y_z})].
\end{equation}
\normalsize
However, from our preliminary experiments, we found that the sigmoid function with cosine similarity score causes numerical problem in optimization. We speculate that the reason for this is that the cosine similarity score (bounded between -1 to +1) only activates the linear regions of the downstream sigmoid activation, reducing model capacity.\footnote{In proxy-triplet loss, this type of numerical problem does not occur because they are relative comparison based losses. In proxy-NCA or proxy-Softmax loss, some of the previous works encountered similar problem, and solved the problem by applying a smoothing factor over the similarity score \cite{qian2019softtriple, zhai2018classification, wu2018unsupervised}. We also tested applying a smoothing factor, but for our multi-label classification problem, it turns out that the proposed modified distance metric is more effective.} Therefore, we modify the distance metric to be

\vspace{-1mm}
\small
\begin{equation} \label{eq:eq6}
D(f(x_i),p_{y_z})=\frac{f(x_i)}{||f(x_i)||}\cdot p_{y_z},
\end{equation}
\normalsize
to ensure that both the learned embedding space is normalized and the sigmoid activations can have nonlinear properties.

From this basic multi-label proxy-based model, we expand the model by applying the masking function as used in the disentangled triplet-based model. Then, the prediction score for each class is updated to 

\vspace{-1mm}
\small
\begin{equation} \label{eq:eq7}
\hat{y_z}=sigmoid(D(f(x_i)\circ m_s,p_{y_z}\circ m_s)),
\end{equation} 
\normalsize
as illustrated in Figure \ref{fig:figure1} (b).

\subsection{Classification-based Model} \label{sec:3.3}
\vspace{-1mm}

Classification-based metric learning has recently been explored  \cite{zhai2018classification, qian2019softtriple}. The core idea is to apply a normalization layer on the embedding feature space. This simple technique ensures that the learned representation  has unit length and makes similarity-based retrieval more effective compared to the vanilla classification model. Therefore, the prediction score of classification-based metric learning model for each class is

\vspace{-1mm}
\small
\begin{equation} \label{eq:eq8}
\hat{y_z}=sigmoid(\frac{f(x_i)}{||f(x_i)||}\cdot c_{y_z}),
\end{equation}
\normalsize
where $c_{y_z}$ is a centroid for each class (parameters of the last hidden layer).\footnote{In our preliminary experiments, we found that removing the bias term does not decrease the model performance, so we did not include it in the Equation \ref{eq:eq8}.} At this stage, we observe that the distance metric inside the sigmoid function of Equation \ref{eq:eq8} is equivalent to that of our modified distance metric in Equation \ref{eq:eq6} of the proxy-based model. 

As for triplet-based metric learning, we extend classification-based metric learning to learn a disentangled embedding space. We begin from the disentangled distance metric, which is  

\vspace{-1mm}
\small
\begin{equation} \label{eq:eq9}
\begin{split}
D(f(x_i)\circ m_s,c_{y_z}\circ m_s) 
=\frac{f(x_i) \circ m_s}{||f(x_i)\circ m_s||} \cdot (c_{y_z} \circ m_s) \\
=\frac{1}{||f(x_i)\circ m_s||}\cdot (f(x_i) \circ m_s)\cdot (m_s\circ c_{y_z}).
\end{split}
\end{equation} 
\normalsize
From the above equation, if we split $f(x_i)$ into the nonlinear function $f_{n-1}(x_i)$ and the embedding feature layer $h$ (here, $h$ layer includes nonlinear activation), then the equation becomes

\vspace{-1mm}
\small
\begin{equation} \label{eq:eq10}
\begin{split}
=\frac{1}{||f(x_i) \circ m_s||}\cdot (f_{n-1}(x_i) \cdot h \circ m_s)\cdot (m_s\circ c_{y_z}) \\
=\frac{1}{||f(x_i) \circ m_s||}\cdot f_{n-1}(x_i) \cdot h \circ m_s\cdot m_s\circ c_{y_z}. 
\end{split}
\end{equation} 
\normalsize
In this equation, $(h \circ m_s\cdot m_s\circ)$ is actually a sub-dense layer that has the same dimensionality as the disjoint mask $m_s$, which is applied when $y_z \in s$. Henceforth, we denote the sub-dense layer $h_s$. Now, $||f(x_i) \circ m_s||$ can be replaced to $||f_{n-1}(x_i) \cdot h_s||$. Finally, the disentangled distance metric becomes 

\vspace{-1mm}
\small
\begin{equation} \label{eq:eq11}
=\frac{1}{||f_{n-1}(x_i) \cdot h_s||}\cdot (f_{n-1}(x_i) \cdot h_s) \cdot c_{y_z}. 
\end{equation} 
\normalsize
This is the same formula for multi-task learning in the multi-label classification problem formulation, surprisingly, proving a previously unknown link between the two concepts. We illustrate this disentangled classification-based model in Figure \ref{fig:figure1} (c). Through experimental evaluation, we further verify that this multi-task learning-based classification model is equivalent to the disentangled proxy-based model while being much simpler to implement and benchmark. 

\section{Experiments}

\subsection{Dataset and Input Features}
\vspace{-1mm}

For our experiments, we use the Million Song Dataset (MSD) \cite{bertin2011million} and Last.FM tag annotations associated with MSD tracks, which have been previously grouped into different categories \cite{choi2017convolutional}, resulting in 28 genre tags, 12 mood tags, 5 instrument tags, and 5 era tags. We treat each category as a similarity notion $s$. We use these tags for evaluating similarity-based retrieval, auto-tagging, and triplet prediction tasks. The data are  split into 201680, 11774, and 28435 samples for the train, validation, and test sets, respectively, following a previous auto-tagging benchmark \cite{lee2017multi}. For triplet prediction evaluation, we follow the same procedure as in \cite{lee2020disentangled}, albeit switch one similarity notion (era replaces tempo) to match auto-tagging benchmarks. We sample 40,000 triplets per each similarity notion (genre, mood, instruments, era, track) and use a cleaned version of the \emph{dim-sim} dataset to evaluate the models on human-annotated triplets.

The input to the embedding function $f(\cdot)$ is 3-second excerpts represented as a log-scaled mel-spectrogram $S$, extracted with librosa \cite{mcfee2015librosa}. We use a window size of 23 ms with 50\% overlap and 128 mel-bands, resulting in input dimensions of $129\times128$ as in \cite{lee2020disentangled}. The input features are z-scored standardized using fixed mean and standard deviation values of 0.2 and 0.25, respectively.

\subsection{Backbone Model and Training Parameters}
\vspace{-1mm}

For the embedding function or backbone model $f(\cdot)$, we use the same architecture as described in~\cite{lee2020disentangled}, which is an Inception-based model~\cite{szegedy2015going}. The model is comprised of a convolution layer with $5\times5$ sized 64 filters followed by $2\times2$ strided max-pooling, followed by six Inception blocks. Each Inception block consist of two Inception modules, a \emph{na\"{i}ve} module and \emph{dimension reduction} module, which are applied in sequence. Both of the modules include filters of mixed size, but the \emph{na\"{i}ve} module has $2\times2$ strides in the last convolution layers of the module, so that the spatial feature map is reduced, and the \emph{dimension reduction} module has a fixed number of filters in the last convolution layers of the module, so that the feature map is fixed to $256$ in the intermediate layers. At the end, one fully connected layer with $256$ units is added, except for the disentangled (multi-task learning) classification-based model, which uses sub-dense layers instead of a single fully connected layer. We use ReLU nonlinearities for all layers.

Since our embedding dimensionality is $256$ and we consider four music similarity notions (genre, mood, instruments, era), each has a disjoint subspace of size $64$. For the disentangled (multi-task learning) classification-based model, the sub-dense layers are also $64$ units each.
We use the Adam optimizer~\cite{kingma2014adam} for training. We initialize the learning rate to $0.005$ and reduce it by a factor of 5 when the validation loss does not decrease for 10 epochs, up to 5 times, after which we apply early stopping. The margin for the triplet-based models is set to $0.1$.

\begin{table*}[t]
\vspace{-2mm}
\centering
\resizebox{0.8\textwidth}{!}{
\begin{tabular}{ccccccccc}
\toprule
\multirow{2}{*}{Models} & \multirow{2}{*}{Normalization} & \multirow{2}{*}{Disentanglement}                          & Training time & \multicolumn{4}{c}{Similarity-based retrieval} & Auto-tagging \\ 
                                               & &  & ratio         & R@1     & R@2     & R@4     & R@8     & AUC          \\ \midrule
Triplet  & \cmark &    \xmark                                     & 1.87          & 31.8    & 45.2    & 59.9    & 73.0    & 0.815           \\
Triplet   & \cmark & \cmark                      & 2.37          & 36.5    & 50.5    & 64.1    & 76.0    & 0.825           \\
Triplet + track reg.  & \cmark & \cmark  & 3.05      & 33.9   & 47.5    &  61.9  & 74.3   & 0.813         \\
Proxy   & \cmark & \xmark                                          & 1.11          & \bf{45.0}    & \bf{58.5}    & \bf{71.0}    & \bf{80.9}    & \bf{0.890}        \\
Proxy  & \cmark & \cmark                         & 1.29          & 44.7    & 58.2    & 70.7    & 80.6    & \bf{0.890}        \\
Classification      & \xmark & \xmark                              & \bf{1.00}          & 6.1     & 11.5    & 21.1    & 35.9    & 0.887        \\
Classification   & \cmark &    \xmark              & \bf{1.00}          & 43.8    & 57.8    & 70.3    & 80.3    & 0.887        \\
Classification  & \cmark & \cmark & 1.27          & 44.7    & 58.4    & 70.7    & \bf{80.9}    & \bf{0.890}        \\ \bottomrule
\end{tabular}
}
\caption{Results for training time, similarity search, and auto-tagging.}
\label{table:table1}
\vspace{-1mm}
\end{table*}

\begin{table}[t]
\vspace{-2mm}
\centering
\resizebox{0.55\columnwidth}{!}{
\begin{tabular}{@{}cc@{}}
\toprule
Model               & AUC   \\ \midrule
CRNN \cite{choi2017convolutional}            & 0.850 \\
Self-attention \cite{won2019toward}      & 0.881 \\
Sample-level ReSE-2 \cite{kim2019comparison} & 0.885 \\
Multi-level \& multi-scale \cite{lee2017multi}         & 0.888 \\
Proposed Model      & \bf{0.890} \\ \bottomrule
\end{tabular}
}
\caption{Auto-tagging SOTA comparison.}
\label{table:table1-2}
\vspace{-1mm}
\end{table}

\subsection{Evaluation Tasks}
\vspace{-1mm}

Our learned representations can be utilized for many applications, so there are many aspects to consider when evaluating representation learning models. Therefore, as a unified evaluation framework, we evaluate the models on four tasks: training time, similarity-based retrieval, auto-tagging, and triplet prediction. 

\subsubsection{Training Time}
\vspace{-1mm}

We first measure the overall training time to see the efficiency of the representation learning model. The training time is calculated as the total number of epochs multiplied by the time consumption of 1 epoch. Then, we report the value as a ratio to the shortest training time.

\subsubsection{Similarity-based Retrieval}
\vspace{-1mm}

For the similarity-based retrieval evaluation, we use the recall@K (R@K) metric to measure retrieval quality following the standard evaluation setting in image retrieval  \cite{oh2016deep,wu2017sampling, movshovitz2017no, qian2019softtriple,zhai2018classification}. This metric is useful for evaluating a search system because it measures the quality of the top K retrieved results, which are more important than long-tail retrieved results. The definition of the standard recall@K that is used for single-label problems is as follows. A query song is used to search a test set of recordings and retrieve similar sounding results. If one of the top K retrieved results has the same class label as the query song, the recall@K is set to 1, otherwise it is set to 0. This process is repeated for all samples in the test set and then averaged.

Our data is multi-labeled, however, so we adapt the standard single-label (multi-class) R@K metric to create a multi-label variant. Our definition is  
\begin{equation} \label{eq:eq12}
R@K =\frac{1}{N}\sum_{q=1}^{N}\frac{n(y^q\cap(\cup_{i=1}^{K}y^i))}{n(y^q)},
\end{equation} 
\normalsize
where $N$ is the number of test samples, $y^q$ is the ground truth labels of a query, and $y^i$ is the ground truth labels of the top K retrieved results. And, $n(\cdot)$ denotes the number of the elements of a set. In this setup, if the set of labels of the top K retrieved results contains all the multiple labels of the query song, the recall@K is set to 1, otherwise it is set to the correct answer ratio.
We report R@K when K is 1, 2, 4, and 8. 

\subsubsection{Auto-tagging}
\vspace{-1mm}

Music auto-tagging has been extensively studied in the literature with diverse model architectures \cite{nam2018deep}. As such, we follow standard benchmarking and evaluation criteria, and report area under the receiver-operator curve (AUC) to measure tag-based retrieval performance.

Unlike the proxy-based and classification-based approaches, the triplet-based model doesn't directly predict a class (or several classes) for a given input. Thus, we use the concept of prototypes to obtain classification result from the triplet-based models \cite{snell2017prototypical}. We first average all the embedding features of the training samples that are assigned to each class label to construct prototype (or centroid) of each class label. Then, we measure a distance between these prototypes and embedding feature of each sample and regard it as a prediction score for classification, which itself is directly used for AUC evaluation.

\subsubsection{Triplet Prediction}
\vspace{-1mm}

Triplet prediction score is simply measured by counting the number of correct predictions among all test triplets. Here, it is regarded as correct if the distance between the embedding features of the anchor and the positive is smaller than that of the distance between the anchor and the negative.

\begin{table*}[t]
\vspace{-2mm}
\centering
\resizebox{0.8\textwidth}{!}{
\begin{tabular}{@{}ccccccccc@{}}
\toprule
Embedding space                      & Models & Normalization & Disentanglement                                          & Genre & Mood & Instruments & Era & Overall \\ \midrule
\multirow{8}{*}{Complete space} & Triplet & \cmark & \xmark                                      & 0.771      & 0.725     & 0.653      &  0.701 &  0.712    \\
                                & Triplet & \cmark & \cmark                        & 0.762      &  0.744    & 0.696      & 0.733  &  0.733  \\
                                & Triplet + track reg.  & \cmark & \cmark   & 0.757      & 0.733    & 0.673   & 0.715 & 0.720  \\
                                & Proxy   & \cmark & \xmark                                            & 0.774      &  0.742   & 0.645     & 0.693  & 0.714    \\
                                & Proxy  & \cmark & \cmark                          &  0.762     & 0.742     &  0.660     & 0.716 & 0.720   \\
                                & Classification   & \xmark & \xmark                                   &  0.783     & 0.745     & 0.659      &  0.723 &   0.728    \\
                                & Classification  & \cmark & \xmark                & 0.776      & 0.747     & 0.647      & 0.704  &  0.719   \\
                                & Classification  & \cmark & \cmark   & 0.758      & 0.742     & 0.659      & 0.715  & 0.719 \\  \midrule
\multirow{4}{*}{Sub-space} & Triplet  & \cmark & \cmark                        &  \textbf{0.790}     & \textbf{0.785}     & \textbf{0.798}      & \textbf{0.797}  & \textbf{0.792}    \\
                                & Triplet track reg.  & \cmark & \cmark   & 0.775      & 0.748    & 0.743 & 0.742 & 0.752    \\
                                & Proxy  & \cmark & \cmark                           &   0.777   &    0.740  & 0.734      & 0.700 & 0.738     \\
                                & Classification & \cmark & \cmark  & 0.775     & 0.739     & 0.732      &  0.701  & 0.737\\ \bottomrule 
\end{tabular}
}
\caption{Results on tag-based triplets.}
\label{table:table2}
\vspace{-2mm}
\end{table*}

\begin{table}[t]
\vspace{-2mm}
\centering
\resizebox{1.0\columnwidth}{!}{
\begin{tabular}{@{}ccccc@{}}
\toprule
Models & Normalization & Disentanglement                                          & Track & Human-labeled  \\ \midrule
Triplet  & \cmark & \xmark                                           & 0.957 & 0.820 \\
Triplet  & \cmark & \cmark                         & 0.964 & 0.820 \\
Triplet + track reg.  & \cmark & \cmark   &  0.961 & \textbf{0.852}  \\
Proxy   & \cmark & \xmark                                            & 0.978 & 0.784 \\
Proxy  & \cmark & \cmark                       & 0.978 & 0.791 \\
Classification   & \xmark & \xmark                                   & 0.978 & 0.780 \\
Classification  & \cmark & \xmark                     & 0.978 & 0.795 \\
Classification  & \cmark & \cmark   & \textbf{0.984} & 0.801 \\ \bottomrule
\end{tabular}
}
\caption{Results on track-based \& human-labeled triplets.}
\label{table:table3}
\vspace{-2mm}
\end{table}

\section{Results}
In Table \ref{table:table1}, we present the results for training time, similarity-based retrieval, and auto-tagging. We compare a total of eight models, which are categorized into three learning methods: triplet-based, proxy-based, and classification-based models.  ``Disentanglement'' indicates whether a CSN masking function is applied to each learning method, and ``Normalization'' indicates whether a normalization layer is applied to the model's embedding layer.
``Track regularization'' (track reg.) indicates whether, in addition to tag-based triplets, we also sample triplets by taking the anchor and positive from the same track and the negative from a different track, as proposed in \cite{lee2020disentangled}. 

First, we see that the training time, represented as the ratio between each model's training time and the training time of the fastest approach, is significantly reduced for the proxy-based and classification-based models compared to the triplet-based models. This is because each training sample for the triplet model is actually composed of 3 inputs (anchor, positive and negative) or even 5 when track regularization is also applied, whereas the proxy-based and classification-based approaches only require one input per training sample.

Second, for similarity-based retrieval, we see that the vanilla classification model without a normalization layer exhibits poor performance. This confirms our conjecture that using the representation learned by the classification model without normalization layer directly is not optimal for similarity-based retrieval, as the model is not optimized based on distances in the learned embedding space. We also see that the proxy- and classification-based models are superior to the triplet-based models across the board. We hypothesize that this is due to the latter strategy using only a single label per training sample, whereas the former two use all \mbox{(multi-)labels} for each training sample, thus exploiting a richer signal during training.

Third, for  auto-tagging, we see that the proxy-based and classification-based models outperform the triplet-based model by a large margin. As expected, the vanilla classification-based model performs well on this task.  In Table \ref{table:table1-2}, we compare our proposed classification-based disentangled model to the state of the art (SOTA) for music auto-tagging. Our model outperforms all baselines, setting the new state-of-the-art for music auto-tagging.

Fourth, for triplet prediction, we report tag-based triplet results in Table \ref{table:table2} using different similarity dimensions (genre, mood, instruments, era), and in Table \ref{table:table3} the results for track-based and human-labeled triplets. The ``Embedding space'' column indicates whether we use the complete embedding space to measure the similarity between pairs of examples, or whether we only use the disjoint sub-space ($f(x_i) \cdot m_s$ or $h_s$) corresponding to the similarity notion $s$ used to sample the test triplets (genre, mood, instruments or era). In  Table \ref{table:table3} we use the complete space.

Fifth, in Table \ref{table:table2} we see that while proxy- and classification-based embeddings are superior for music retrieval and tagging, triplet-based embeddings perform better (unsurprisingly) on the triplet-prediction task.
It is noteworthy that while the triplet task is often used as a proxy for evaluating music similarity modelling, models that do best on this task are not necessarily the best at downstream retrieval tasks as evidenced by Table \ref{table:table1}.
In Table \ref{table:table3}, we also see that while classification-based embeddings perform better at predicting track-based triplet similarity, triplet-based embeddings perform better when it comes to matching human judgements of triplet similarity. This is particularly true when we apply triplet learning with track regularization, in accordance with previous work \cite{lee2020disentangled}. 
%

\section{Visualization of Disentangled Space}

\begin{figure}[t!]
\vspace{-2mm}
\centering\includegraphics[width=\columnwidth]{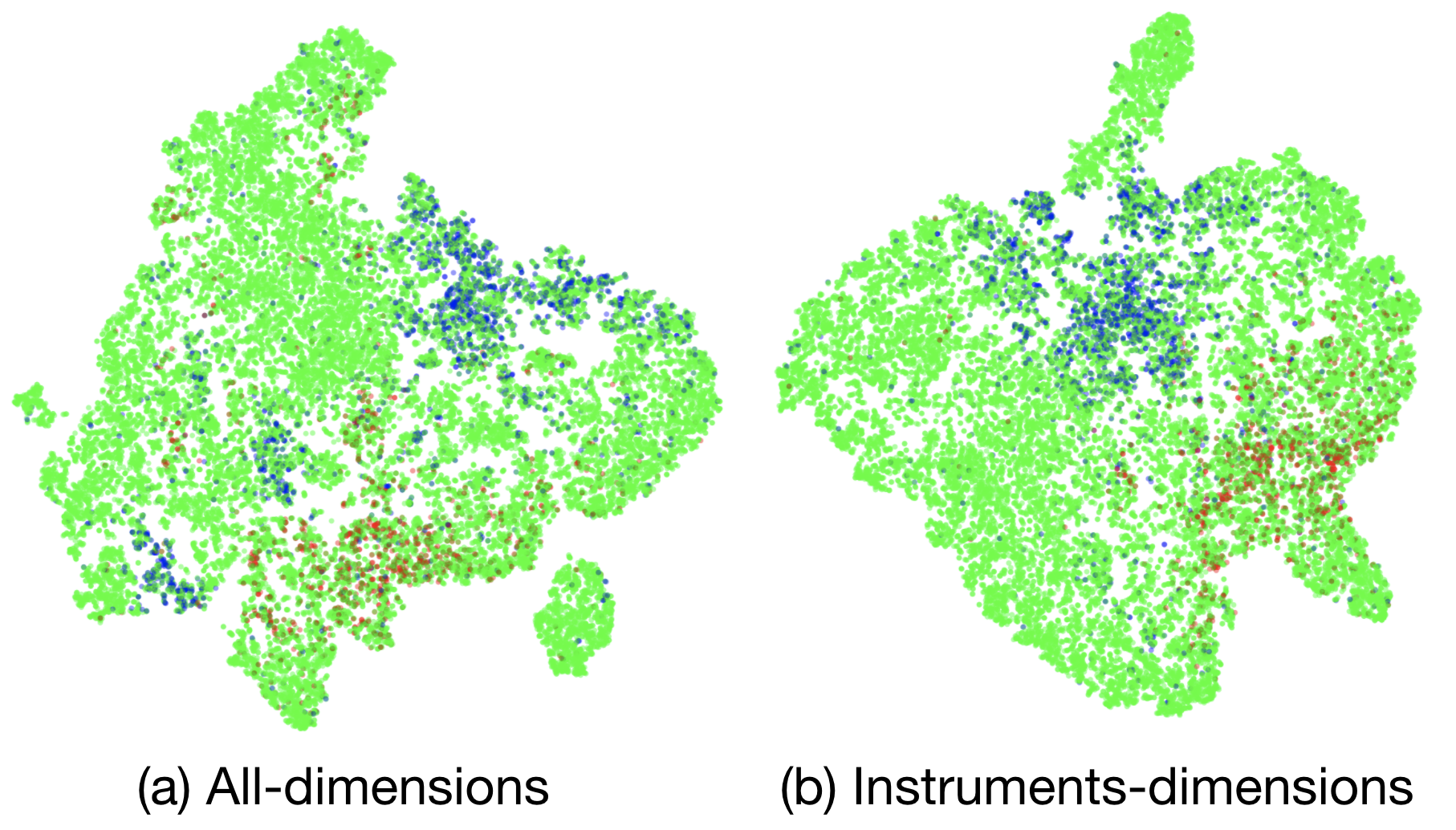}
\caption{t-SNE plot of test set embedding features. The blue dots are labeled positive for the female vocalists tag, the red dots are labeled positive for the instrumental tag, and the green dots are negative.}
\label{fig:figure2}
\vspace{-1mm}
\end{figure}

To qualitatively evaluate the disentangled representation space learned by our model, we visualize the embeddings of the test set as a t-SNE plot \cite{maaten2008visualizing} in Figure \ref{fig:figure2}. We take embeddings from the disentangled triplet model and highlight samples with the \emph{female vocalists} and \emph{instrumental} tags as an example. 
While the highlighted samples are relatively dispersed when considering all dimensions, we see that they are nicely clustered together when only considering the instrument sub-space of the embedding. This illustrates the benefits of a disentangled space, which supports both global similarity and specialized similarity over specific music dimensions.

\vspace{-2mm}
\section{Conclusion}
In this paper, we presented a detailed study of metric-based and classification-based learning approaches for music representation learning. We extended both strategies to learn disentangled spaces from multi-label data, and showed both analytically and empirically that under certain conditions, proxy-based learning is equivalent to classification-based learning. We benchmark multiple variants of each strategy in terms of training efficiency and performance on music retrieval, auto-tagging, and triplet prediction tasks. Our results show that, when coupled with disentanglement and normalization, classification-based representation learning produces superior benchmark results on all tasks, except for triplet prediction where triplet models are (predictably) strong performers, indicating that triplet prediction is not necessarily a reliable proxy for real-world retrieval performance. Our best performing disentangled model obtains state-of-the-art results for music auto-tagging, outperforming all previous baselines. Finally, we complement our quantitative analysis with qualitative results that further illustrate the benefits of learning a disentangled music embedding space.

\bibliography{ISMIRtemplate}

%
%
%
%

\end{document}